\documentclass[12pt]{iopart}

\usepackage[squaren,Gray]{SIunits}
\usepackage{graphicx}
\usepackage{subfigure}
\usepackage{color,soul}
\usepackage[normalem]{ulem}
\usepackage[sort]{cite}

\newcommand{\mnras}{Mon. Not. R. Astron. Soc.}       
\newcommand{\nat}{Nature}

\newcommand{\pasp}{Publ. Astron. Soc. Pacific}

\newcommand{\eul}{Europhysics Letters}
\newcommand{\epjd}{European Physical Journal D}
\newcommand{\fpp}{Foundations of Probability and Physics - 3}
\newcommand{\rmps}{Reviews of Modern Physics Supplement}

\newcommand{\NIMPA}{Nuclear Instruments and Methods in Physics Research A}

\newcommand{\njp}{New J. Phys.}

\def\gsim{\lower.73ex\hbox{$\sim$}\llap{\raise.4ex\hbox{$>$}}$\,$}
\def\lsim{\lower.73ex\hbox{$\sim$}\llap{\raise.4ex\hbox{$<$}}$\,$}
\def\ggsim{\lower.73ex\hbox{$\sim$}\llap{\raise.4ex\hbox{$\gg$}}$\,$}

\begin{document}

\title{A proposal for measuring photon temporal coherence in continuum radiation}

\author{Richard Fong
}

\address{Department of Physics, University of Durham, Rochester Building, Science Laboratories, South Road, Durham DH1 3LE, UK}
\ead{richard.fong@durham.ac.uk}

\begin{abstract}
A technique complementary to those for spectral lines is proposed for the observation of continuum radiation. As, quantum mechanically, the radiation is a mixture of pure states, it should be possible to measure the temporal coherence of the states as a function of energy. We propose here experiments with just this possibility, showing the empirical grounds on which it is based. A simple means of measuring temporal coherence is with the extent of the interference pattern produced in a Young's double-slit experiment. Thus, we propose the use of such an experiment, but with a screen of energy measuring photon detectors, small enough and sensitive enough to delineate the interference patterns according to photon energy. We then only register a detection if the absorbed photon is found to have the energy of interest. In this way, we have allowed the interference, i.e., the wave aspect of the experiment, to proceed as normal, recovering the interference pattern by accumulating just those registered detections. We, also, explain why this form of filtering does not actually violate the conventional wisdom that, when using an optical filter with incident continuum radiation to separate out the photons of a particular energy, only the temporal coherence of the filter's passband would be measured. It naturally then introduces the concept of a detections filter, differentiating it from an optical filter.
\end{abstract}

\vspace{2pc}

\ 

\submitto{\njp}

\ 



\maketitle

\section{Introduction}
\label{intro}
The measurement of spectral lines is an important investigative tool. From our personal perspective in astronomy, together with photometry, they constitute the principal observational tools, with polarimetry, perhaps, of more specialised use. For cosmology, the measurement of the redshift, $z$, has been absolutely crucial. It has led to our present understanding that we inhabit a Big Bang universe. The recent discovery that  matter is dominated by exotic dark matter is an exciting one and points towards a particle physics beyond the standard model. A further revelation is that our Universe is, instead of decelerating, actually accelerating. It has led to a universe in which, at the present epoch, as much as 70\% of the total energy is dark energy, which is simplest interpreted as vacuum energy. If this is confirmed to be equivalent to Einstein's cosmological constant, then, under the Big Bang scenario, it is an energy density that is constant back to the time of inflation and would, thus, be a relic of the Big Bang itself, from when the Universe was younger than $\sim 10^{-32}$ secs.

On the other hand, in the case of one of the principal cornerstones of cosmology, the cosmic microwave background does not have any spectral lines, possessing to remarkable accuracy a Planck spectrum. Its redshift is then based on its observed temperature relative to that at recombination. However, without this physical insight, it would be difficult to establish its redshift, since when redshifted or blueshifted any thermal radiation remains Planckian in form. There are, also, occasions when there is only one spectral line observable, risking a misidentification. Worse, of course, is when a distant source simply displays a continuum spectrum with no significant spectral features at all.

Now, just as the wavelength of a photon from a distant source is increased by $(1+z)$, so is its temporal coherence. Thus, it is interesting that, quantum mechanically, continuum radiation is seen as a mixture of photons, more rigorously, as a mixture of pure states. It is with this in mind that we considered that a technique for estimating the temporal coherence of photons in continuum radiation could be similarly as useful as that of spectral line measurements. However, since the physics is not as simple as that for the central wavelength of a spectral line, such a novel technique may prove more powerful for understanding the physical processes of the source and, as such, it would also be useful in the laboratory. Indeed, for fundamental physics, it would further promise a new empirical means of exploring the nature of continuum radiation.

Our proposal then is for an experimental technique to measure in the optical the temporal coherence of a subset of states in the continuum radiation, a subset defined by photon energy. Normally, to select out such a subset, an optical filter would be used. But, in that case, there is the conventional wisdom that with incident continuum radiation it is only possible to measure the temporal coherence of the passband of the spectral filtering apparatus used. It would seem then that such a proposal as ours is, unfortunately, not possible. Consequently, we devote a completely separate brief section, Sec.~\ref{conundrum}, to addressing this conundrum.

In the main section of our paper, Sec.~\ref{empbasis}, we present the empirical basis for a slightly modified double-slit experiment for measuring the photon temporal coherence. This is followed in Sec.~\ref{constraints} with a consideration of the experimental constraints. We have just mentioned Sec.~\ref{conundrum}. Sec.~\ref{popexp} proposes an actual initial experiment, opening the way to developing the technique for its present use in both astronomy and the laboratory. Finally, we present in Sec.~\ref{conclusions} some brief conclusions.

\section{The experimental basis}
\label{empbasis}
A simple means of estimating the temporal coherence is through the extent of the interference as seen in the pattern produced when the light is incident on a Young's double-slit experiment, Fig.~\ref{fig:exp}. As the wave aspect of our proposal is no different from that for a normal double-slit experiment, there is no need for a more complex treatment than can be found in standard textbooks on the subject. In such treatments, however, the radiation is treated completely classically, whereas here we wish to exploit our quantum mechanical insight into continuum radiation as being a mixture of pure states.

Basic to our proposal is Dirac's famous remark\cite{Dirac1958}: ``Each photon interferes \ldots only with itself. Interference between two different photons never occurs''. It's a statement that we now see as not applying generally, but, happily, does apply to experiments generic to the Young's double-slit experiment\cite{Glauber1965,Scully1997}. Although it is consistent with Taylor's early experiment with extremely feeble light\cite{Taylor1909}, it has had to wait until the recent single photon interference experiments\cite{Grangier1986,Jacques2005,Aichele2005,Mandel1999,Zeilinger2005} to be firmly established experimentally, requiring as it does the development of sources of single isolated photons. Of course, in our case we do not have single isolated photons incident; we could actually have an innumerable number incident, as, for example, in the case of one source of interest, that of the sun. But, importantly, it is these single photon interference experiments that provide the first of the two pieces of empirical evidence underpinning our proposal. They imply that, for experiments such as ours, we can treat each photon separately and simply accumulate the observations photon by photon to obtain the resulting interference pattern\cite{Zeilinger2005}. It's most graphically illustrated by images such as the ones in Jacques et~al.~(2005)\cite{Jacques2005}, where it is shown in their Figs.~4(a)--(c) how the individual detections do, in fact, build up to form the interference pattern.

Focusing then on an individual photon, we can describe it, physically, as a wave packet of the electromagnetic field incident upon the detection screen, depicting in Fig.~\ref{fig:exp} an idealised photon as a wave train of wavelength, $\lambda_{s}$, and length, $l_s$, and, so, a photon with energy, $E_{s} = h c / \lambda_{s}$, and a temporal coherence of 
\begin{equation}
\tau_s=l_s/c.      \label{eq:tcohln}
\end{equation}
We consider then our continuum source as consisting of a mixture of such photons over a continuous range of wavelengths. With the usual setup, the photons are detected using a screen of small photon detectors, such as the CCDs in current use, or, as in the early experiments, using a photographic plate as the detector screen. Thus, when a photon reaches the screen and is absorbed, a detector is triggered and the detector's position, $y$, is registered, graphically, as a detection on the screen.

\setcounter{figure}{0}

\begin{figure}
  \center
  \includegraphics[width=10cm]{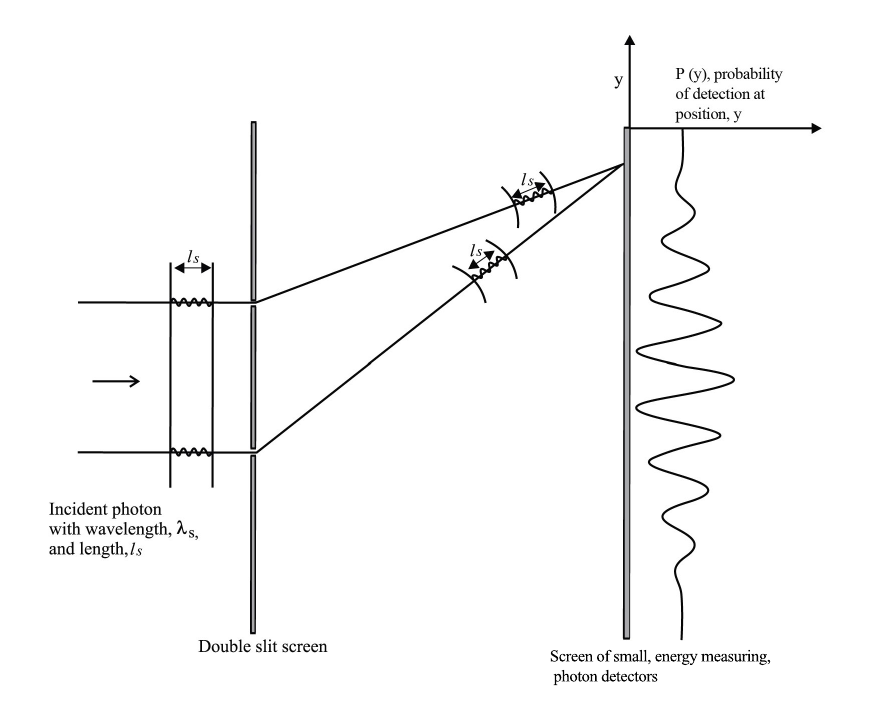}
  \caption{{\bf Proposed double-slit interference experiment.} 
The figure shows a typical setup for a Young's double-slit experiment, however, with a modest requirement as to the form of the detection screen. Depicted, then, is a screen made up of small photon detectors capable of measuring the energy deposited by a photon in the detector.\label{fig:exp}}
\end{figure}

Returning to the idealised photon in Fig.~\ref{fig:exp}, it then has the probability, $P(y)$, of triggering a detector at $y$, where $P(y)$ is determined in the usual way by the form of the wave train. Thus, if we consider incident just photons all with the same wave train, the single photon interference experiments shows that the interference pattern formed by accumulating these detections on the detection screen would have the form of $P(y)$. Consequently, we shall simply also refer to $P(y)$ as the interference pattern. As is well known, when the path difference of the two paths in Fig.~\ref{fig:exp} exceed the coherence length, $l_s$, there is no interference and $P(y)= {\mathrm{constant}}$. Thus, by examining where the fringes disappear in the interference pattern, the experiment provides a measure of the photon temporal coherence, since the number, $n_{\mathrm{f}}$, of fringes in the interference pattern is simply related to $l_s=c\tau_s$:
\begin{equation}
n_{\mathrm{f}}= 2 {l_{s} \over \lambda_{s}}.      \label{eq:fringeno}
\end{equation}
Of course, the physics here applies generally to any incident state, including multi-photon states, such as those produced in stimulated emission, as the important measure here is just the path difference for which any interference is seen to disappear. But, the idealised wave packet depicted in Fig.~\ref{fig:exp} does lend itself more easily to simple mathematical expressions, so, serving well the pedagogical aim of this section.

Now, when continuum radiation is incident, we, of course, just obtain an interference pattern of white light fringes and the point where the fringing disappears for any individual wavelength is lost. However, the interference pattern for just those photons with energy, $E_{s}$, is still present. All that it requires is to identify, among the complete set of photon detections, just those on the detection screen for which the absorbed photon has this particular energy of $E_{s}$.

A widely held view, however, is that in such experiments the ``final state of the field is never measured''\cite{Scully1997}. Although Glauber makes the same statement in his 1964 Les Houches lectures\cite{Glauber1965}, it is qualified by a further remark that in ``the detectors used to date \ldots the final states form \ldots a continuum'', keeping open the possibility of future technology changing the circumstance. In fact, he does also remark in passing on the possible use of atoms with discrete final states, something that we shall return to in Sec.~\ref{popexp}. For the present, there is, indeed, vigorous ongoing activity to measure the final state energy; it is, of course, just a calorimetric measurement of the absorption process. It provides the second piece of experimental evidence supporting our proposal, that the energy deposited by a photon on a CCD pixel can, in principle, be measured. In particular, it has been demonstrated that the energy of an X-ray photon absorbed by a CCD can be accurately measured\cite{Lumb1990}; it's a procedure that is routinely carried out in X-ray astronomy to obtain the X-ray spectrum of a distant source. In the optical, there is also ongoing research into energy measuring photon detectors, such as with superconducting tunnel junctions and similar low temperature photon detectors\cite{Perryman1993,Peacock1996,Jakobsen1999,Mather1999,Cropper2003,Mazin2013}. However, the challenge for our proposal is to find or develop detectors for the optical band that can not only measure the energy of the absorbed photon, but are also small enough to delineate with sufficient clarity the interference pattern.

So, given such photon detectors, by registering only the detections for those photons with energy, $E_{s}$, the interference pattern, $P(y)$, formed by these detections is just that for those states in the mixture having a photon with energy, $E_{s}$. Measuring the extent of the interference seen in the pattern provides then the desired estimate of the temporal coherence for this subset of states. Of course, repeating the experiment for different values of $E_{s}$ provides this photon temporal coherence as a function of energy.

But, clearly, all that we have done here is to put forward the case that the experimental evidence for our proposal is, actually, already established in the results of the single photon interference experiments. All that remains is just to implement a means of also measuring the energy of the absorbed photon. It may also be worth noting how the physical discussion here recalls the semi-classical treatment of the Hanbury~Brown--Twiss effect\cite{Hanbury1956}, in that, we treat the photon as a classical wave for the interference part of the experiment, but then apply quantum physics for the detection process\cite{Fox2006}. It emphasises our quantum understanding that the phenomena of interference is a consequence of the wave aspect of photons, with the particle aspect governing the detection process. We need, nevertheless, to remember the quantum mechanical nature of the initial premise behind our proposal, the insight into continuum radiation as a mixture of pure states and in the fact that a photon only interferes with itself, at least, for experiments such as ours.

\section{Experimental constraints}
\label{constraints}
For the pedagogical aim of the previous section, we used an idealised picture of a photon. In actuality, the important element is the state, with continuum radiation  a mixture of pure states. The observed temporal coherence would then be some average over the states in the mixture that can give rise to a photon detection with an energy of $E_{ s}$. Indeed, in representing a photon as a wave packet, we see that a photon does not even possess a definite energy, since the Fourier transform of a finite wave packet consists of Fourier components over a range of wavelengths and, thus, a range, albeit generally a small range, of energies. In fact, just through the uncertainty principle, a finite temporal coherence necessarily implies a finite energy width. Thus, there will also be contributions from states for which the most probable energy would not be $E_{ s}$, although it would generally be close. As it is rather awkward to write of a state able to excite a photon detector through the absorption of a photon of energy, $E_{ s}$, we shall simply talk, albeit rather loosely, of the state as having a photon of this energy and of the observed temporal coherence obtained in this way as the photon temporal coherence.

Also, there can, in reality, be no exact measurement of the energy deposited by a photon. We would certainly expect significant thermal effects for the photon detectors. So, in reality, the observed temporal coherence is, actually, an average over the states with photons in the energy range, $E_{s}\pm \Delta E_{d}$, where $\Delta E_{d}$ is the energy resolution of the detectors.

Clearly, for any meaningful measure of photon temporal coherence, the detector bandwidth needs to be kept as narrow as possible, ideally, with
\begin{equation}
\Delta\nu_{d} \ll \Delta\nu_{s},        \label{eq:constraint}
\end{equation}
where $\Delta\nu_{d}=\Delta E_{d}/h$ is the frequency bandwidth of the detectors and $\Delta\nu_{s}=1/\tau_s$, with $\tau_s$, then, an `average' temporal coherence for this subset of states in the mixture that have a photon in the energy range, $E_{s}\pm \Delta E_{d}$. This is in direct contrast to the normal use of broadband detectors, where the relation is usually the other way round with $\Delta\nu_{d} \gg \Delta\nu_{s}$, such as, for example, with how CCDs are presently used in the optical. Physically, the constraint of Eq.~\ref{eq:constraint} is simply stating that, for a measurement of the photon temporal coherence, the observable zone for interference needs to be clearly greater than the actual area of interference.

\section{The experiment as apparent counter-example}
\label{conundrum}
Basic to our proposal is the ability to separate the results due to the subset of states with a photon with energy, $E_{s}$, from the combined results due to all the states in the mixture that is the continuum radiation. And the obvious means for doing so would be to make use of an optical filter. Unfortunately, with optical filters, there is the conventional wisdom, that with continuum radiation incident the measured temporal coherence would be just that for the bandwidth of the filter, showing that our proposal would, in that case, be unachievable. However, although our proposed energy measuring photon detectors are acting as an effective narrow band filter, the act of filtering in our case is radically different from that of a normal optical filter.

In our case, the filtering is done by simply selecting from the complete set of photon detections just those with a measured energy of $E_{s}$, more rigorously, $E_{s}\pm \Delta E_{d}$, with, as discussed in Sec.~\ref{empbasis}, the pattern formed by their positions just the interference pattern for this particular energy. Thus, there is no filtering of the light. Indeed, it is extremely important that the wave aspect of the experiment is untouched, which would not be the case for an optical filter. Thus, our experiment here is not even a counter-example to the conventional wisdom. To emphasise this, we propose to call such a detection screen a detections filter, differentiating it from a normal optical filter.

\section{Proposal for an initial experiment}
\label{popexp}
We now recognise that in restricting the photon detections to those for which the photon's energy has been found to be $E_{s}$ we have simply turned the detection screen into one of effective two level atoms, at least, as far as the absorption process is concerned, with $E_{s}$ as the energy level difference. However, basing our discussion in Sec.~\ref{empbasis} on the detectors used in the single photon interference experiments, such as the CCD detectors of Jacques et~al.~(2005)\cite{Jacques2005}, allowed us to show that the experiment we are proposing is already embedded in their results. Unfortunately, to implement the proposal as set out in Sec.~\ref{empbasis} would require photon detectors that, for the moment, are only a future prospect. But, with the insight we now have, it suggests the possible use of photon detectors with discrete final states, something Glauber himself had alluded to\cite{Glauber1965}. Simplest, we think, would be to use photoluminescent atoms/molecules. Actually, this also recalls the use of a phosphor screen for a realisation of Feynman's double-slit thought-experiment for demonstrating the wave character of electrons\cite{Bach2013}.

The photoluminescent particles may either be embedded in some transparent material or layered on a dark surface. Their importance here lies in the fact that, with the finite time delay between absorption and emission, the incident photon has, clearly, been absorbed in the same way as in a normal experiment, where a screen of broadband CCDs is used or, for that matter, a photographic plate. In particular, it is important for us that the excitation of a photoluminescent particle constitutes the particle detection of an incident photon. The observed interference pattern can then only be due to the photons that have the right energy to excite a photoluminescent particle into its higher energy level. Thus, the measured temporal coherence, as obtained from the extent of the interference, would be for this subset of photons in the incident continuum radiation, i.e., for just those photons having an energy in the range, $E_{s}\pm\Delta E_{d}$, where $E_{s}$ is the energy difference between the unexcited and excited levels of the photoluminescent particles and $\Delta E_{d}$ is the width of the energy range over which a transition can occur. Furthermore, if necessary, one could also cool the detection screen to bring $\Delta E_{d}$ closer to the natural width of the transition.

For the test source, the continuum radiation from a hot dense gas or plasma should prove particularly effective. The photons' temporal coherence from such a source would be dominated by collisional broadening. Provided the constraint of Eq.~\ref{eq:constraint} is met, this would be observable in the resulting interference pattern. The test is further strengthened by increasing the pressure whilst keeping the source at constant temperature, whereby fewer fringes should then be seen due to the shorter temporal coherence caused by the increased rate of collisions in the gas/plasma. Also, of course, there will be a need to compare the observations with theory. Alternatively, if one were available, a more appealing source might be that of a clock-triggered photon source. Although not a continuum source, it would provide a demonstration of the need to satisfy Eq.~\ref{eq:constraint}. And, it does have the advantage of greater control over a source's photon temporal coherence.

\section{Conclusions}
\label{conclusions}
The motivation behind our proposal has been with the quantum mechanical insight into continuum radiation as a mixture of pure states. For discussing the physical basis of our proposal, we considered a Young's double-slit experiment, but with a screen of small photon detectors that can also measure the energy of the absorbed photon. The photon temporal coherence is then given by the extent of the interference in the pattern formed by the positions of just those detections of a photon with the particular energy of interest, $E_{s}$. We showed in Sec.~\ref{empbasis} the empirical basis for such measurements and, in particular, how its potential is, actually, already apparent in the single photon interference experiments\cite{Grangier1986,Jacques2005,Aichele2005,Mandel1999,Zeilinger2005}.

Using a detection screen with the ability of selecting out just detections of a particular energy is, of course, a form of filtering. But, it is a filtering process that is radically different from that of an optical filter. With an optical filter, the filter is placed in the path of the radiation, crucially, altering the photon's wave function before it undergoes absorption by the photon detector. Whereas with our setup the radiation is allowed to continue to the detection screen as normal and the filtering done through registering, after photon detection, only the detections for the particular energy of interest. To differentiate it from normal optical filters, we have, thus, called this form of filtering a detections filter.

As we have pointed out, the development of the technique we propose here would be a welcome additional observational tool for astronomy. For example, in the case of solar radiation, a coherence time of $\sim10^{-8}$~secs is generally expected. Interestingly, Smid (2006)\cite{Smid2006} proposes a model in which the photons in the sun's radiation is calculated to have a coherence time of $\sim10^{-12}$~secs. Clearly, this has implications for the mass density at the base of the photosphere. For a more distant prospect, the temporal coherence of the photons in the cosmic microwave background would be exciting, especially since they originate from electron-positron annihilation during the `annihilation era', when the Universe was still just $\sim10$~secs old. It could even be used to estimate the relative redshifts between two sources if it was thought that the physics of both sources were the same or, perhaps more essentially, to test this for similar sources at the same redshift. Generally, the technique holds out the promise of a new and rich vein of astronomical observations for exploring the physics of distant sources.

It would, clearly, also be a useful complementary tool to those already in use in the laboratory and in industry, both as one for testing theoretical calculations, as well as its use as a diagnostic tool. For fundamental physics, it would provide a new means of experimentally examining the nature of continuum radiation. Interestingly, the observation of plasmas and gases in the laboratory would be useful for astronomy too, not only in providing confidence in theoretical calculations, but, even more directly, when there is the possibility of replicating an astronomical source in the laboratory. Finally, we note that we have just concentrated here on the photon temporal coherence. Obviously, there will be further information about the source in the visibility of these energy selected interference patterns, i.e., in the form of their envelopes.

\ack

It is a pleasure to acknowledge discussions with many colleagues in the Physics Department here. In particular, Tom Shanks has acted, with much good humour, as a constant sounding board for the author, despite it not being in his actual area of expertise. Those with Robert Potvliege also helped clarify the essential quantum nature of the proposal. Thanks also to Thomas Smid for  his patience with the author, who is no authority on plasma physics! Michael Perryman and Gordon Love for bringing to the author's attention superconducting tunnel junctions and similar detectors. The basic idea motivating the paper was mooted a quarter of a century ago! Unfortunately, the conventional wisdom explained the experiment the author had proposed at the time. The author would then like to acknowledge those long ago discussions with Roy Pike and his team for humouring him, also encouraging replies from Rod Davies, Daniel James and the late Leonard Mandel. For the present paper, Daniel James and John Girkin have been kind enough to read through it and make useful comments, helping clarify the paper. Also, grateful acknowledgement of the helpful comments from referees.

\newpage
 
\section*{References}
\bibliographystyle{unsrt}

\begin{thebibliography}{1}

\bibitem{Dirac1958}
P.~A.~M.~Dirac 1958
   \newblock {\em The Principles of Quantum Mechanics} 4th edn
\newblock (Oxford University Press)

\bibitem{Glauber1965}
R.~J.~Glauber 1965
 \newblock Optical Coherence and Photon Statistics.
 \newblock In {\em Quantum Optics and Electronics}
 \newblock (eds C.~Dewitt, A.~Blandin and C.~Cohen-Tannoudji) 63--185 (Gordon and Breach)
 
 \bibitem{Scully1997}
M.~O.~Scully and M.~S.~Zubairy 1997
 \newblock {\em Quantum Optics}
 \newblock (Cambridge University Press)
 
 \bibitem{Taylor1909}
G.~I.~Taylor 1909
 \newblock Interference fringes with feeble light
 \newblock {\em Proc. Camb. Phil. Soc.} {\bf 15} 114--115
 
\bibitem{Grangier1986}
P.~Grangier, G.~Roger and A.~Aspect 1986
 \newblock Experimental Evidence for a Photon Anticorrelation Effect on
a Beam Splitter: A New Light on Single-Photon Interferences
 \newblock {\em \eul} {\bf 1} 173--179
 
\bibitem{Jacques2005}
V.~Jacques, E.~Wu, T.~Toury, {\em et~al.} 2005
 \newblock Single-photon wavefront-splitting interference
 \newblock {\emph \epjd} {\bf 35} 561--565

\bibitem{Aichele2005}
T.~Aichele, U.~Herzog, M.~Scholz and O.~Benson 2005
 \newblock Single-photon generation and simultaneous observation of
wave and particle properties
 \newblock {\emph \fpp} {\bf 750} 35--41

\bibitem{Mandel1999}
L.~Mandel 1999
 \newblock Quantum effects in one-photon and two-photon interference
 \newblock {\em \rmps} {\bf 71} 274--282

\bibitem{Zeilinger2005}
A.~Zeilinger, G.~Weihs, T.~Jennewein and M.~Aspelmeyer 2005
 \newblock Happy centenary, photon
 \newblock {\em \nat} {\bf 433} 230--238
 
\bibitem{Lumb1990}
D.~H.~Lumb 1990
 \newblock Applications of charge coupled devices to X-ray astrophysics missions
 \newblock {\em \NIMPA} {\bf 288} 219--226
 
\bibitem{Perryman1993}
M.~A.~C~ Perryman, C.~L.~Foden and A.~Peacock 1993
 \newblock Optical photon counting using superconducting tunnel junctions
  \newblock {\em \NIMPA} {\bf 325} 319--325

\bibitem{Peacock1996}
A.~Peacock, P.~Verhoeve, N.~Rando,~et~al. 1996
 \newblock Single optical photon detection with a superconducting tunnel junction
  \newblock {\em \nat} {\bf 381} 135--137
 
\bibitem{Jakobsen1999}
P.~Jakobsen 1999
 \newblock Superconducting Tunnel Junction Detectors for Optical and UV Astronomy.
 \newblock In {\em Ultraviolet-Optical Space Astronomy Beyond HST}
 \newblock (eds J.~A.~Morse, J.~M.~Shull and A.~L.~Kinney) 397--404 (ASP Conference Series 164)
 
 \bibitem{Mather1999}
 J.~C.~Mather 1999
 \newblock Astronomy: Super photon counters 
   \newblock {\em\nat} {\bf 401} 654--655
   
\bibitem{Cropper2003} M.~Cropper, M.~Barlow, M.~A.~C.~Perryman,~et~al. 2003
 \newblock A concept for a superconducting tunnelling junction based spectrograph
   \newblock {\em \mnras} {\bf 344} 33--44
   
\bibitem{Mazin2013} B.~A~Mazin, S.~R.~Meeker, M.~J.~Strader,~et~al. 2013
 \newblock ARCONS: A 2024 Pixel Optical through Near-IR Cryogenic Imaging Spectrophotometer
   \newblock {\em \pasp} {\bf 125} 1348--1361

\bibitem{Hanbury1956}
R.~H.~Brown and R.~Q.~Twiss 1956
 \newblock Correlation between Photons in two Coherent Beams of Light
 \newblock {\em \nat} {\bf 177} 27--29

\bibitem{Fox2006}
M.~Fox 2006
 \newblock {\em Quantum Optics: An Introduction}
 \newblock (Oxford University Press)
 
 \bibitem{Bach2013}
 R.~Bach, D.~Pope, S.-H.~Liou and H.~Batelaan 2013
 \newblock Controlled double-slit electron diffraction
   \newblock {\em \njp} {\bf 15} 033018

\bibitem{Smid2006}
T.~Smid  2006
 \newblock Photoionization Theory for Coherent and Incoherent Light
  \newblock  $<$http://www.plasmaphysics.org.uk/photoionization.htm$>$

\end{thebibliography}

\end{document}